\documentclass[12pt,nohyper,letterpaper]{JHEP3}         
\usepackage{amssymb,amsfonts}

\usepackage{epsfig}
\usepackage{amsmath}

\def\be{\begin{eqnarray}}
\def\ee{\end{eqnarray}}
\newcommand{\nn}{\nonumber}
\newcommand\para{\paragraph{}}
\newcommand{\ft}[2]{{\textstyle\frac{#1}{#2}}}
\newcommand{\eqn}[1]{(\ref{#1})}

\def\Dslash{\,\,{\raise.15ex\hbox{/}\mkern-12mu D}}
\def\Zslash{\,\,{\raise.15ex\hbox{/}\mkern-12mu Z}}
\def\Wslash{\,\,{\raise.15ex\hbox{/}\mkern-12mu W}}
\def\zslash{\,\,{\raise.12ex\hbox{/}\mkern-10mu z}}
\def\wslash{\,\,{\raise.12ex\hbox{/}\mkern-13mu w}}
\def\Dbarslash{\,\,{\raise.15ex\hbox{/}\mkern-12mu {\bar D}}}
\def\delslash{\,\,{\raise.15ex\hbox{/}\mkern-9mu \partial}}
\def\delbarslash{\,\,{\raise.15ex\hbox{/}\mkern-9mu {\bar\partial}}}
\def\pslash{\,\,{\raise.15ex\hbox{/}\mkern-9mu p}}
\def\calDslash{\,\,{\raise.15ex\hbox{/}\mkern-12mu {\cal D}}}

\newcommand{\e}{\,{\rm e}}

\newcommand{\RN}{Reissner-Nordstr\"om\ }

\newcommand{\pp}{\langle \bar{\psi}\psi\rangle}
\newcommand{\pfivep}{\langle \bar{\psi}\gamma^5\psi\rangle}
\newcommand{\plrp}{\langle {\bar{\psi}}_L \psi_R\rangle}
\newcommand{\PP}{\langle \bar{\Psi}\Psi\rangle}

\def\lae{\mathrel{\mathop{\smash{\lower .5 ex \hbox{$\stackrel<\sim$}}}}}
\def\lae{\mathrel{\mathop{\smash{\lower .5 ex \hbox{$\stackrel>\sim$}}}}}

\def\theequation{\thesection.\arabic{equation}}

\title{Magnetic Catalysis in AdS${}_4$}
\author{
Stefano Bolognesi${}^{1,2}$ and David Tong${}^{1}$ \\
${}^1$Department of Applied Mathematics and Theoretical Physics, \\
University of Cambridge, UK\\
${}^2$Racah Institute of Physics, The Hebrew University of Jerusalem, 91904, Israel
{\tt s.bolognesi, d.tong@damtp.cam.ac.uk}
}

\abstract{We study the formation of fermion condensates in Anti de Sitter space. 
In particular, we describe a novel version of magnetic catalysis that arises for fermions in asymptotically AdS${}_4$ geometries which cap off in the infra-red with a hard wall.  We show that in the presence of a magnetic field, a $\pfivep$ condensate develops in the bulk, spontaneously breaking $CP$ symmetry.  From the perspective of the dual boundary theory, this corresponds to a strongly coupled version of $d=2+1$ magnetic catalysis.}

\begin{document}
\pagestyle{plain} \setcounter{page}{1}
\newcounter{bean}
\baselineskip16pt

\section{Introduction}

Chiral symmetry breaking usually occurs only when the interaction between fermions is sufficiently strong and the   coupling exceeds some critical value. But, in the presence of a magnetic field, things work differently. By trapping particles in the lowest Landau level, a background magnetic field effectively reduces the dimensionality of the system, promoting infra-red effects. In many cases, this drives the critical coupling to zero so that arbitrarily weak attractive interactions will result in a condensate for massless fermions, together with the associated chiral symmetry breaking.  This phenomenon is known as {\it magnetic catalysis} \cite{klim,miransky1,miransky2,miransky3}. 

\para
Perhaps the simplest example of magnetic catalysis occurs in $d=2+1$ dimensions.  Here the discussion is usually framed in terms of a pair of two-component Dirac spinors $\Psi_1$ and $\Psi_2$.  The ``chiral" symmetry is  a $U(2)$ flavour symmetry which is broken when a parity preserving mass $m$ is introduced. In the presence of a background magnetic field one indeed finds that arbitrarily weak interactions are sufficient to form  a condensate. Importantly, this survives even in the massless limit where the condensate spontaneously breaks chiral symmetry,
\be \lim_{m\rightarrow 0^+}\, \langle\bar{\Psi}_1\Psi_1- \bar{\Psi}_2\Psi_2\rangle = -\frac{B}{2\pi}\label{intro1}\ee
This form of magnetic catalysis (albeit with four Dirac fermions rather than two) has been invoked \cite{khves,mir4,mir5,herbut} to explain the quantum Hall plateaux in graphene which emerge  at filling fractions $\nu=0$ and $\nu=\pm 1$ in very strong magnetic fields \cite{zhang}.

\para
There is a similar story with just a single Dirac fermion in $d=2+1$ dimensions. In this case, a mass term $m$ necessarily breaks both parity, $P$, and parity combined with charge conjugation, $CP$. Once again, in the presence of an external magnetic field, the condensate survives in the limit of vanishing mass
\be  \lim_{m\rightarrow 0^+}\, \langle\bar{\Psi}\Psi\rangle = -\frac{B}{4\pi}\label{intro2}\ee
This time there is no continuous chiral symmetry to break and the magnetic field itself breaks $P$. However, the discrete symmetry $CP$ is spontaneously broken by the condensate. (We will be more specific about how the different objects above transforms under various discrete symmetries later in this introduction).

\para
The condensates above are derived at weak coupling. Indeed, on dimensional grounds,  equations \eqn{intro1} and \eqn{intro2} can only hold at weak coupling. The dimension of the external magnetic field is always $[B]=2$ while only a free fermion has dimension $[\Psi]=1$. We could ask what becomes of these condensates in strong coupling situations where  $\Psi$ develops a large anomalous dimension. Does the condensate still survive in the massless $m\rightarrow 0$ limit? In this case, the right-hand side of the \eqn{intro1} and \eqn{intro2} must necessarily become non-analytic in $B$.  Or do these condensates vanish in a such strongly interacting systems?

\para
One of the purposes of this paper is to compute the condensate in a class of strongly interacting theories in which the dimension $[\Psi]$ is far from its free value. However, along the way we will also see a novel and interesting phenomenon of magnetic catalysis in four dimensions. The full story is somewhat intricate and we will use the remainder of this introduction to summarize the main points.

\subsection{Summary}

In this paper, we tell two intertwined stories of magnetic catalysis. The first story revolves around Dirac fermions  in $d=3+1$ dimensional anti-de Sitter space. The second story is the holographic image of the first, projected onto the $d=2+1$ dimensional boundary.

\para
In AdS${}_4$, our protaganist is a massless bulk Dirac fermion $\psi$. The narrative is as follows:
\begin{itemize}
\item For massless fermions in AdS, it's not necessary to turn on a magnetic field to induce a $\pp$ condensate. Such a condensate occurs even in pure AdS and arises because the bulk chiral symmetry is necessarily broken by boundary conditions. At one-loop, these boundary conditions infect the bulk, resulting in a constant $\pp$ throughout AdS.
\item In the presence of a magnetic field, the bulk geometry changes to the \RN black hole. The condensate $\pp$ now picks up a radial profile, but otherwise is little changed. In particular, no symmetries are spontaneously broken.
\item More interesting dynamics occur when an IR cut-off is introduced in the AdS geometry. This can be most simply implemented through a hard wall construction. Now massless particles can bounce backwards and forwards between the UV boundary and the IR wall. In this set-up, a novel form of magnetic catalysis occurs: the presence of a background magnetic field induces the condensate $\pfivep$. 
\end{itemize}

\para
The story above is told in Section 2. The most interesting aspect is the formation of the $\pfivep$  condensate. This occurs due to infra-red divergences in the bulk which arise as the fermion undergoes an infinite number of bounces between the boundary and hard wall. These IR divergences result in a  non-analytic term in the fermion propagator which, in turn, give rise to the condensate.

\para
The formation of a $\pfivep$ condensate spontaneously breaks $CP$ symmetry in the bulk. To see this, recall that the magnetic field $B$ is odd under parity (Larmor orbits go in the opposite direction in the mirror) and odd under charge conjugation (electrons and positrons go in different directions in a magnetic field) and therefore even under $CP$. The transformation properties of $B$, the four-dimensional condensates $\pp$ and $\pfivep$, and the three dimensional condensate $\langle\bar{\Psi}\Psi\rangle$, are tabulated below.

\begin{center}
\begin{tabular}{c|cccc}  & $\pp_{\rm 4d}$ & $\pfivep_{\rm 4d}$ & $\langle\bar{\Psi}\Psi\rangle_{\rm 3d}$ &  B  \\ \hline 
$P$ & + & - & - & - \\
$C$ & + & + & + & - \\ $CP$ &  + & - & - & +
\end{tabular}
\end{center}


\para
Our second story of magnetic catalysis concerns the implications of this bulk physics for the dual boundary theory. The bulk spinor $\psi$ is dual to a boundary fermion $\Psi$ which is  a two component Dirac spinor operator in $d=2+1$. Our main result in Section 3 is to identify the bulk and boundary operators
\be \bar{\psi}\gamma^5\psi \ \longleftrightarrow \ \bar{\Psi}\Psi  \label{intro3}\ee
This mapping is to be understood in the usual manner of AdS/CFT; a constant expectation value for $\bar{\psi}\gamma^5\psi$ corresponds to a source for  $\bar{\Psi}\Psi$; the sub-leading terms encode the condensate expectation value  $\langle\bar{\Psi}\Psi\rangle$. Notice, however,  that this goes beyond the usual AdS/CFT dictionary. It matches a multi-particle state in the bulk to a double trace operator in the boundary. The formation of the bulk condensate is a one-loop effect while the existence of the boundary condensate is a ``1/N" correction to the statement that all gauge invariant correlation functions factorize at large N. 

\para
One simple, but compelling, piece of evidence for the relationship \eqn{intro3} is the matching of the discrete symmetries presented in the table above. We give further evidence in Section 3.  The upshot is that magnetic catalysis does not happen in the simplest conformal field theories with an AdS dual; however, when the IR is cut-off --- in our case, with a the hard wall --- it does take place.

\para
There have been previous discussions of magnetic catalysis in a holographic framework, although all with a somewhat different perspective \cite{filev1,filev2,filev3,filev4,nick}. (There has also been discussion of  an inverse magnetic catalysis effect in the Sakai-Sugimoto model \cite{inverse}). In all these papers, one studies a bulk scalar field dual to a single trace fermion bilinear, $\phi \sim {\rm Tr}(\bar{\lambda}\lambda)$, which is identified with the embedding coordinate of a probe D-brane. In this setting, magnetic catalysis of a single trace operator is identified with the bending of the brane in a background magnetic field. As should be clear from the discussion above, we take an alternative approach, instead mapping magnetic catalysis of a bulk fermion to that of a double trace operator $\bar{\Psi}\Psi$.

\section{Fermion Condensates in the Bulk}\label{secbulk}

We consider a single, four-component Dirac spinor $\psi$ propagating in AdS${}_4$. We work in the Poincar\'e patch of AdS. We will primarily use the radial coordinate $r$ such that the boundary sits at $r\rightarrow \infty$, with the metric given by
\be ds^2 = \frac{r^2}{L^2}(-dt^2+dx^2+dy^2)+\frac{L^2}{r^2}dr^2\nn\ee
However, at times it will prove more convenient to use the reciprocal coordinate
\be z = \frac{L^2}{r}\nn\ee
for which the boundary is at $z=0$. 

\para
The bulk fermion dynamics is governed by the action
\be S_{\rm bulk}  = \int d^4x \sqrt{-g}\ \left[\frac{i}{2}\bar{\psi}e^\mu_a\,(\gamma^a \!\stackrel{\rightarrow}{D}_\mu-\stackrel{\leftarrow}{D}_\mu\!\gamma^a)\psi 
-im\bar{\psi}\psi\right]\label{sbulk}\ee
where $e^{\mu}_a$ is the vielbein and the covariant derivative is $ D_{\mu} = \partial_{\mu} + \omega_{ab\mu} \gamma^{ab}/4-iA_\mu$. Here  $\omega_{ab \mu}$ is the spin connection and $A_\mu$ is a background gauge field which we  will later turn on to give a magnetic field. We use the basis of gamma matrices
\be \gamma^0 = \left(\begin{array}{cc} 0 & i\sigma^3 \\ i\sigma^3 & 0 \end{array}\right)\ \  ,\ \ 
\gamma^i = \left(\begin{array}{cc} 0 & \sigma^i \\ \sigma^i & 0 \end{array}\right),\  i =1,2\ \ , \ \
\gamma^r = \left(\begin{array}{cc} 1 & 0 \\ 0 & -1 \end{array}\right) \ \  , \ \ 
\gamma^5 = \left(\begin{array}{cc} 0 & i \\ -i & 0 \end{array}\right)\nn\ee
Due to the presence of the AdS boundary, in order to have a well defined variational principle, we must augment the bulk action with the boundary term. We choose
\be S_{\rm boundary} = \frac{i}{2}\int_{r\rightarrow \infty} d^3x\sqrt{gg^{rr}}\ \bar{\psi}\psi\label{sboundary}\ee
The requirement that $\delta S_{\rm bulk} + \delta S_{\rm boundary} = 0$ holds if the spinor obeys the bulk Dirac equation, together with boundary condition 
\be \psi_+ \equiv \frac{1}{2}(1+\gamma^r)\psi = 0\ \ \ \ {\rm at}\ r=\infty\ \label{psiplus}\ee
This choice of boundary condition is usually referred to  as ``standard quantization" (even though it has little to do with quantization -- it is needed to define the classical bulk dynamics).

\subsection{Chiral Symmetry in AdS}\label{chiralsec}

Throughout this paper, we work with a  massless fermion, setting $m=0$. (This is primarily for calculational convenience. We expect  that our results will carry over in spirit to the massive case and it would be interesting to confirm this). There are a number of rather special properties of the massless fermion that can be traced to the chiral symmetry enjoyed by the bulk action \eqn{sbulk},
\be \psi \rightarrow \e^{i\alpha\gamma^5/2}\psi\label{chiral}\ee
%
\EPSFIGURE{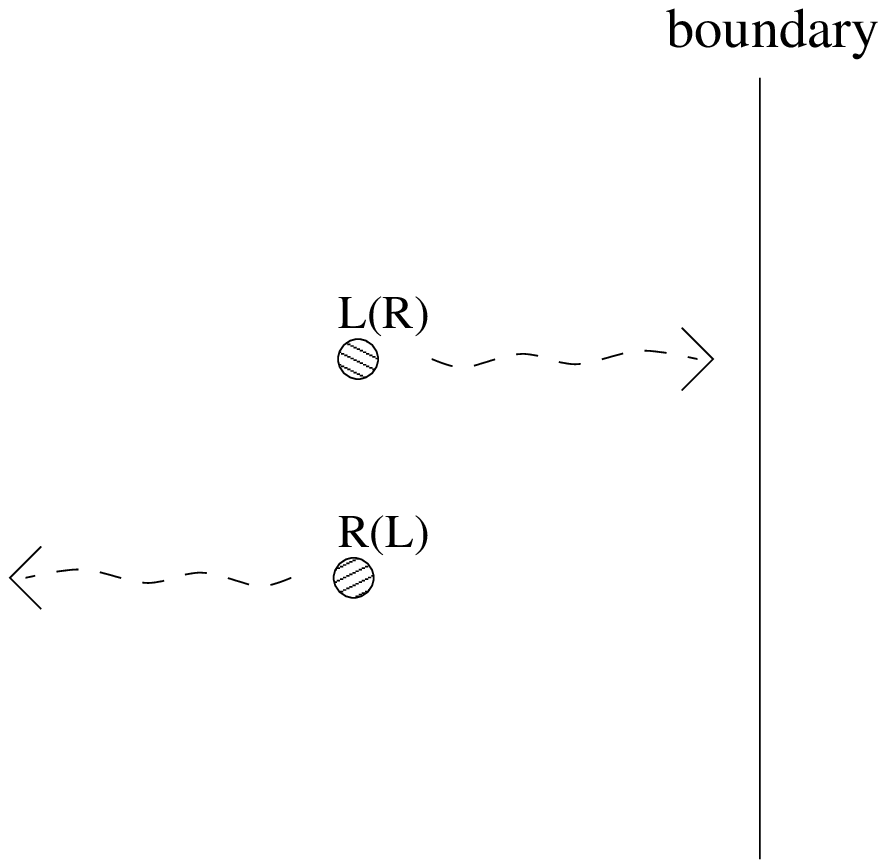,height=120pt}{Flipping Helicity}
The chiral symmetry is broken by the boundary condition \eqn{psiplus}. Perhaps the most physical implication of this chiral symmetry breaking is seen by thinking in terms of particles. It is well known that a massless particle reaches the boundary of AdS in finite time before bouncing back into the bulk.  The boundary conditions \eqn{sboundary} ensure that a particle moving purely in the radial direction will reflect off the wall with its momentum reversed and its spin unchanged. Or, in other words, its helicity is reversed: a right-handed particle returns left-handed. 

\para
Arguments of this kind --- what turns into what after bouncing off what --- will be increasingly important. For this reason, it's useful to spell this out in more detail why the AdS boundary inverts the helicity as claimed. To do so, we  look  at the solutions to the massless, bulk  Dirac equation. It is convenient to do this using  the $z=L^2/r$ coordinates, where the boundary is at $z=0$.  A typical wavefunction obeying the bulk Dirac equation is then given by
\be \psi = \left(\frac{z}{L^2}\right)^{3/2} \,\left[e^{i\omega(t+z)}\left(\tiny{\begin{array}{c} 1 \\ 0 \\ -i \\ 0 \end{array}}\right) - e^{i\omega(t-z)}\left(\tiny{\begin{array}{c} 1 \\ 0 \\ i \\ 0 \end{array}}\right)\right]\nn\ee
The first term corresponds to a right-handed particle moving towards the AdS boundary; it has eigenvalue $+1$ under $\gamma^5$. The second term is needed to ensure that the boundary conditions \eqn{psiplus} are obeyed. It corresponds to a left-handed particle moving away from the boundary.

\para
There is one further implication of chiral symmetry that will be important for us: this is a circle of possible boundary conditions, first described in \cite{porrati,rr}. This circle arises by acting on the boundary term \eqn{sboundary} with the chiral symmetry \eqn{chiral}. This gives rise to a new boundary term,
\be S_{\rm boundary} = \frac{i}{2}\int_{r\rightarrow \infty} d^3x\sqrt{gg^{rr}}\ \bar{\psi}e^{i\alpha\gamma^5}\psi\label{newbound}\ee
The point $\alpha=0$ reduces to the boundary condition \eqn{sboundary}; the point $\alpha=\pi$ corresponds to what is usually called ``alternative quantization". For $m=0$, these two points are continuously connected by the one-parameter family of boundary conditions above. In the dual boundary theory, this corresponds to the existence of a line of large $N$ fixed points, induced by the marginal double trace operator $\bar{\Psi}\Psi$ \cite{wittem,micha}. We will have more to say about this in Section \ref{secbound}. 

\para
Re-doing the analysis above, we learn that the general $\alpha$ boundary condition has the following effect on ingoing particles: a right-handed fermion returns left-handed, but now also picks up a phase $e^{i\alpha}$; a left-handed fermion returns right-handed, now with a phase $e^{-i\alpha}$. 

\para
This circle of boundary conditions will be very important for us when computing the effects of magnetic catalysis. To prepare for this, it will be useful to introduce the familiar notation for left- and right-handed spinors
\be \psi_L = \frac{1}{2}(1-\gamma^5)\psi \ \ \ ,\ \ \ \psi_R=\frac{1}{2}(1+\gamma^5)\psi\label{leftright}\ee
The advantage of this is that one can introduce a single complex fermion bi-linear,
\be \bar{\psi}_L\psi_R= \frac{1}{2}\bar{\psi}\psi + \frac{1}{2}\bar{\psi}\gamma^5\psi\nn\ee
It is easy to disentangle the contribution from $\bar{\psi}\gamma^5\psi$ and $\bar{\psi}\psi$ since the former is real, while the latter is purely imaginary\footnote{Note that $\pp$ is imaginary because, in signature $(-+++)$, $\gamma^0$ is anti-hermitian.}. This means that we can extract the condensates $\pfivep$ and $\pp$ that we are interested in from the magnitude and phase of $\langle \bar{\psi}_L\psi_R\rangle$. 

\subsection{Boundary Contamination: $\langle\bar{\psi}\psi\rangle$}\label{rrsec}

The phenomenon of magnetic catalysis describes the formation of chiral symmetry breaking condensates, such as $\pp$, due to an external magnetic field. But in AdS, you don't need a magnetic field to induce a fermion condensate: the condensate appears already in vacuum as a consequence of the boundary conditions. This effect was described long ago in \cite{allen} and discussed more recently in the context of AdS/CFT in \cite{grip,rr}. Here we  provide a simple derivation of the condensate. 

\para
We work with a massless fermion $m=0$ and start with the standard boundary condition \eqn{sboundary}. The condensate is a one-loop effect and can be thought of as an infection: the breaking of chiral symmetry by the boundary leaks into the bulk, manifesting itself through the presence of a $\pp$ condensate. 
To see this, we employ the form of the spinor bulk-to-bulk propagator given in \cite{kawano}. 
%
%
%
%
%
%
%
%
%

\para
For a fermion of general mass $m$, the propagator in \cite{kawano} involves sums of hypergeometric functions.  But for $m=0$, something nice happens. These hypergeometric functions reduce to elementary functions and the propagator becomes much less intimidating.
It is once again convenient to use the reciprocal coordinate $z=L^2/r$ and we denote two points in the bulk by $Z=(z,\vec{x})$ and $W=(w,\vec{y})$.  The massless propagator is 
\be S(Z,W) = -\frac{i}{2\pi^2}\,\left(\frac{zw}{L^2}\right)^{3/2}\,\left[\frac{\Zslash-\Wslash}{[(z-w)^2 + (\vec{x}-\vec{y})^2]^2} - \frac{\Zslash\gamma^r+\gamma^r\Wslash}{[(z+w)^2 + (\vec{x}-\vec{y})^2]^2} \right]\ \ \ 
\label{sprop}\ee
These two terms have a very simple interpretation which follows from the fact that a massless fermion in AdS is conformally related to a massless fermion in flat space with a boundary. The first term represents direct propagation from $Z$ to $W$; the second term represents propagation from $Z$ to the mirror image of $W$ which lies behind the boundary at $(-w,\vec{y})$. The conformal factor, $(zw)^{3/2}$, can be seen sitting out front. The presence of the $\gamma^r$ factors in the second term reflects the helicity-violating boundary condition \eqn{psiplus}.

\para
The condensate is easily computed by taking $Z\rightarrow W$. The divergence in the first term of \eqn{sprop} is the usual divergence in flat space and can be dealt with by simply taking the trace over the gamma matrices from the beginning\footnote{For a massive fermion, taking the trace is not sufficient to remove the divergence and one must subtract the flat space divergence from the AdS result.}. We're then left with the condensate
\be \pp  = \lim_{Z\rightarrow W}\, {\rm Tr}\,S(Z,W) =  \frac{i}{4\pi^2L^3}\nn\ee
This is reminiscent of the mechanism of generating chiral symmetry breaking through confinement in the bag model of QCD \cite{casher}. Here AdS acts like a (very big) bag.

\para
It is  simple to generalise the computation of the condensate to a generic point $\alpha$ on the circle of possible boundary conditions. Indeed, the answer follows simply by acting with the chiral transformation \eqn{chiral} which ensures that the phase of the  $\langle \bar{\psi}_L\psi_R\rangle$ condensate follows the phase of the boundary term, 
\be \plrp = \frac{ie^{i\alpha}}{8\pi^2L^3}\label{plrp}\ee

\subsection{Turning on the Magnetic Field}\label{nobhsec}

This paper is primarily concerned with fermionic condensates in the presence of a magnetic field. We will go slowly, introducing one new piece at a time until we build the full picture.  

\para
The key bit of physics which underlies all magnetic catalysis phenomena is the effective dimensional reduction due to the magnetic field. Semi-classically, particles are trapped in Larmor orbits; quantum mechanically they are restricted to Landau levels. In strong magnetic fields, only the lowest Landau level is important and the dynamics is reduced from $d$ spatial dimensions to $d-2$. 

\para
Let us flesh out this dimensional reduction in more detail for fermions in AdS with a background magnetic field $B$ in the radial direction. For our purposes, it will suffice to keep the background metric as AdS, neglecting the backreaction of the magnetic field on the geometry. (The reason for this will become clear in the next section). 
In the appendix, we provide a computation in the \RN black hole background.
We'll start by considering a $d=2+1$ dimensional slice of the AdS${}_4$ geometry at fixed $r$. We choose the gauge 
\be A_x=By\nn\ee
Ignoring the radial direction for the time being, we decompose the four-component spinor $\psi$ into two two-component spinors $\chi_1$ and $\chi_2$ with the ansatz,
\be \psi = e^{-i\omega t + ikx}\left(\begin{array}{c}\chi_1(y) \\ \chi_2(y)\end{array}\right)\label{itsthis}\ee
The massless Dirac equation reduces to a 3d Dirac equation for each of the $\chi_i$,
\be \left(\begin{array}{cc}\omega & i(k-By)-i\partial_y \\ i(k-By)+i\partial_y & -\omega\end{array}\right) \chi_i=0\nn\ee
It is straightforward to reduce this equation to that of the simple harmonic oscillator. The eigenvalues are given by
\be \omega = \sqrt{2Bn}\ \ \ \ \ n=0,1,2,\ldots\nn\ee
\FIGURE{\epsfig{file=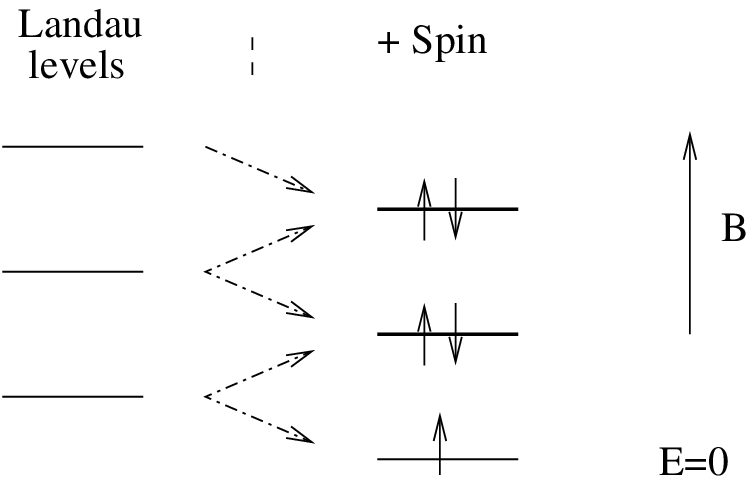,width=5cm} 
        \caption[]{Zeeman Splitting of the Landau Levels}%
	\label{zeromodes}}
\noindent These, of course, are the Landau levels. Importantly, the lowest Landau level for massless relativistic fermions  has vanishing energy. This is because the usual $+\ft 12 B$ zero point energy of the lowest Landau level is precisely compensated by the $-\ft 12 B$ shift in energy which arises from the spin coupling. This Zeeman splitting is shown in Figure \ref{zeromodes}.  

\para
The zero modes for each of the two component spinors  $\chi_1$ and $\chi_2$ are given by
\be \chi_i =  e^{-B(y-kB)^2/2}\left(\begin{array}{c} 0 \\ \xi_i\end{array}\right)\nn\ee
The fact that the zero mode has support only in the second component of $\chi_i$ is telling us that the lowest Landau level is spin polarised. In contrast, all higher Landau levels include states with both spins as shown in the figure. This means that the density of states of the lowest Landau level is $B/2\pi$, while all higher Landau levels have density $B/\pi$. 

\para
With our interest in low energy physics, we focus only on the lowest Landau level. We now re-introduce the radial direction $r$, promoting the fermionic zero modes to functions $\xi_i(r,t)$. There is one such zero mode for each state of the lowest Landau level, i.e. for each value of $k$.  Each of these zero modes is governed by a $d=1+1$ dimensional action for the 
two-component spinor $\xi_i = (\xi_1,\xi_2)$, which arises upon substitution into the action\footnote{Strictly speaking, the action \eqn{seff} only arises from \eqn{sbulk} when the fermionic zero modes are appropriately normalised. In particular, this means taking a suitable Fourier sum over $k$ modes. The choice of normalisation, while changing intermediate steps,  does not affect the end result for $\pp$ that we're interested in.} \eqn{sbulk},
\be S_{\rm eff} = \int dtdr\ \sqrt{-g}\ i\bar{\xi}\!\Dslash\xi \label{seff}\ee
In this expression, the  two-dimensional gamma matrices implicit in $\Dslash$ are given by $\gamma^\mu_{\rm 2d} = (-i\sigma^1,\sigma^3)$.  The covariant derivative now contains the spin connection, but no gauge field. 

\para
The computation of the condensate is entirely analogous to the calculation in Section \ref{rrsec}. The 2d propagator arising from \eqn{seff} is once again conformal to the flat space propagator for a fermion on the half line. There are two terms: 
one arising from direct propagation; the other from a bounce. After taking the trace, only the bounce survives. It is most useful to express the resulting propagator in terms of 2d chiral fermions  $\xi_L$ and $\xi_R$, the definition of which can be traced to the 4d definition \eqn{leftright}.
\be \xi_L = \frac{1}{2}(1-\sigma^2)\xi \ \ \ ,\ \ \ \xi_R=\frac{1}{2}(1+\sigma^2)\xi\nn\ee
We work with general $\alpha$ boundary condition \eqn{newbound}. Inserting the fermions at equal times, but different radial positions, the contribution to the propagator from the bounce is 
\be \bar{\xi}_L(z)\xi_R(w) = -\frac{ie^{i\alpha}}{4\pi}\,\left(\frac{zw}{L^2}\right)^{3/2}\frac{\zslash\sigma^3+\sigma^3 \wslash}{(z+w)^2}\nn\ee
where the overall factor of $e^{i\alpha}$ is the phase picked up by a right-handed fermion after bouncing off the boundary, as  discussed  in Section \ref{chiralsec}. The $(zw)^{3/2}$ term is the conformal factor and the remaining term is familiar as the flat space propagator. We can immediately compute the condensate of the two-dimensional fermion by taking the trace and sending $z\rightarrow w$,
\be \langle\bar{\xi}_L\xi_R\rangle = -\frac{ie^{i\alpha}}{4\pi} \frac{z^2}{L^3} \nn\ee
To compute the  $\psi$ condensate, we must sum over all fermi zero modes\footnote{From \eqn{itsthis}, we see that any individual zero mode is not translationally invariant in the $(x,y)$ plane. One restores translational symmetry only upon integrating over all modes labelled by $k$.} . This just gives a factor of the density of states which, in lowest Landau level, is $B/2\pi$,
\be \plrp = \frac{B}{2\pi} \langle\bar{\xi}_L\xi_R\rangle \label{2d4d}\ee
Finally, reverting to our original coordinates $r=L^2/z$, we learn that the contribution to the condensate from the lowest Landau level is
\be \plrp
=  - \frac{ie^{i\alpha}BL}{8\pi^2 r^2}\label{monday}\ee
It's worth making an obvious point about this result: nothing particularly surprising happens with the phase $e^{i\alpha}$ of the condensate. Just as in pure AdS, the phase  follows that of  the boundary. In particular, if $\alpha=0$, we have a $\pp$ condensate, but $\pfivep=0$. (This ``obvious" statement will cease to be true in the example described in the next section).

\para
The condensate \eqn{monday} arises only from the lowest Landau level.  In flat space, the gap to the next level is $\sqrt{B}$, but in AdS this is warped to $\sqrt{B}L/r$. As we approach the boundary, the gap goes to zero which is to be expected since the magnetic field is becoming dilute there. (Of course, $B$ itself stays constant, but the metric factor enlarges). So for processes that occur close to the boundary, we expect the higher Landau levels to play an important role, resulting in the constant condensate \eqn{plrp} that we previously computed. The cross over from the constant to $1/r^2$  behaviour happens at  $r\sim \sqrt{B}L^2$.

\subsection{Between a Boundary and a Hard Place: $\langle \bar{\psi}\gamma^5\psi\rangle$}\label{hardwallsec}

In this section, we introduce the feature that makes things more interesting. We impose a second, reflecting boundary condition on the fermions in the infra-red. For now, we impose this in the most brutal fashion possible; a hard wall at $r=r_\star$ where AdS ends. Such hard wall geometries have long been used in the study of AdS/QCD \cite{josh} as a way to avoid the complexities of more honest geometries that exhibit confinement \cite{witten,klebstrass}. More recently, it was noted that placing fermions at finite density in such a hard wall geometry leads a holographic construction of a Landau fermi liquid \cite{sachdev}. Our goal here is to study magnetic catalysis in this system.

\para
With a hard wall at $r=r_\star$, we must also impose IR boundary conditions on the fermion. Just as in the  UV, there is a one-parameter family of possible boundary conditions that are related by a chiral transformation. We label them by $\alpha^\prime$, 
\be S_{\rm wall} = -\frac{i}{2}\int_{r=r_\star} d^3x\sqrt{gg^{rr}}\ \bar{\psi}e^{i\alpha^\prime\gamma^5}\psi\label{wallbound}\ee
The overall minus sign arises because this term cancels contributions from the lower limit of the bulk integral \eqn{sbulk} rather than the upper end.
Only the relative phase between the wall and the UV boundary is important. For this reason, we set $\alpha^\prime =0$ but keep the possibility of a general phase $\alpha$ in the UV boundary condition \eqn{newbound}.

 \FIGURE{\epsfig{file=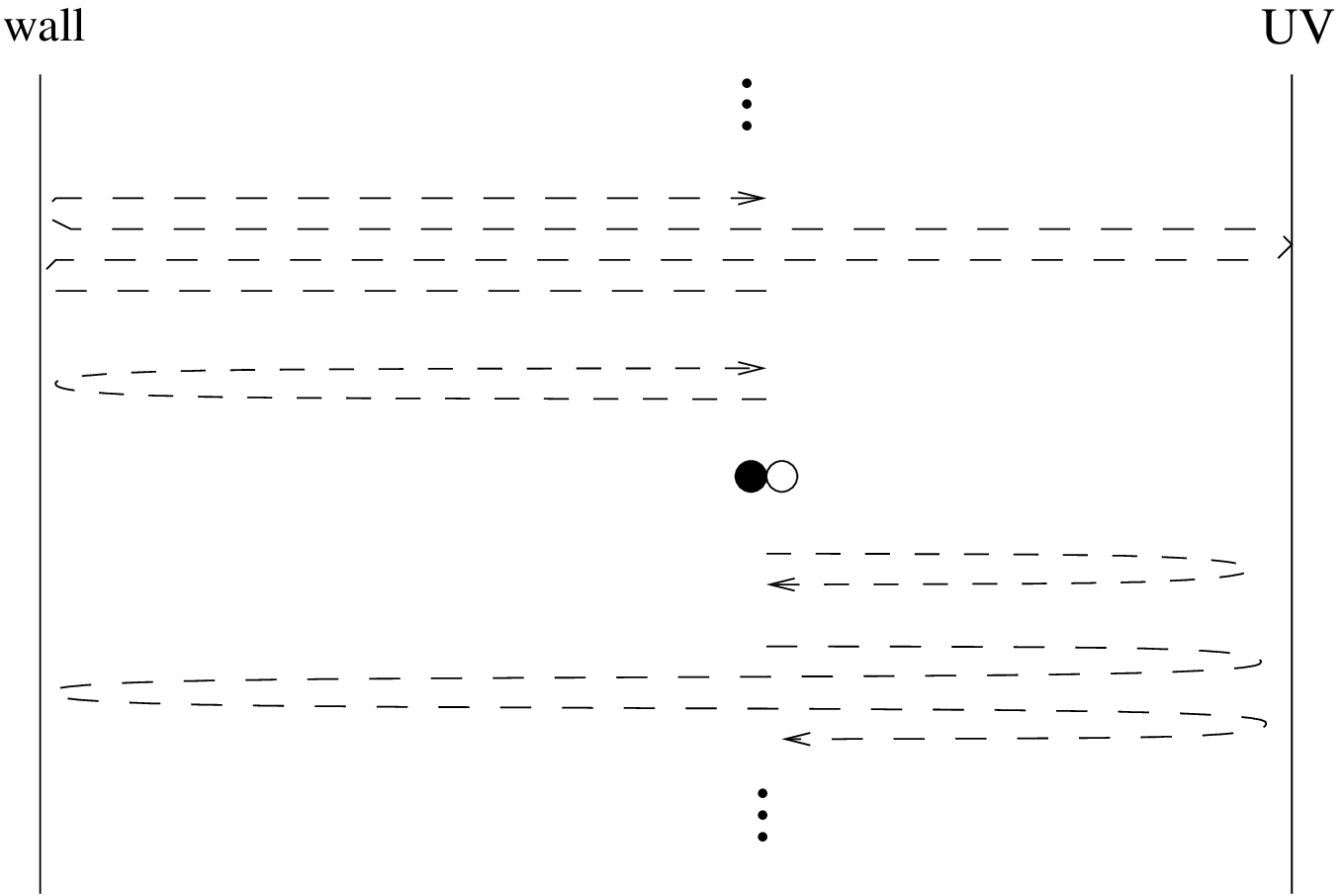,width=6cm} 
        \caption[]{Many bounces}\label{multibounces}}
\para
With two walls for the fermions to bounce off, we have more contributions to the propagator in which the a particle  traverses the distance between the two walls many times.  Every time the fermion bounces off either boundary, its chirality is flipped. In addition, it picks up a phase $e^{i\alpha}$ when bouncing off the UV boundary. 

\para
To compute $\langle\bar{\xi}_L\xi_R\rangle$, only the trajectories with an odd number of bounces contribute since chirality is not flipped after an even number of bounces.  It is useful to distinguish between trajectories that start their life moving towards the UV boundary and those which start moving towards the IR hard wall. 

\para
As discussed in Section \ref{chiralsec}, right-handed fermions which begin by moving towards  the UV wall pick up a phase $e^{i\alpha}$ for each bounce. This gives rise to the series
\be \langle\bar{\xi}_L\xi_R\rangle_{\rm That\ Way} &=& -\frac{i}{2\pi} \left(\frac{z}{L}\right)^3\left[ \frac{e^{i\alpha}}{2z} +\frac{e^{2i\alpha}}{2z+2z_\star} + \frac{e^{3i\alpha}}{2(z+2z_\star)} + \ldots \right] \nn\\ &=& -\frac{ie^{i\alpha}}{4\pi} \left(\frac{z}{L}\right)^3 \sum_{n=0}^\infty \frac{e^{in\alpha}}{z+nz_\star}\nn\ee
where we have again flipped to coordinates $z = L^2/r \in [0,z_\star]$. Each contribution to the condensate picks up an extra factor of $e^{i\alpha}$, reflecting the bounce from the UV wall, while the denominator increases by $2z_\star$, twice the distance between the walls. 

\para
Right-handed fermions which start life by moving towards the IR hard wall are left-handed by the time they bounce off the UV. They pick up a phase $e^{-i\alpha}$ for each bounce, resulting in the series
\be \langle\bar{\xi}_L\xi_R\rangle_{\rm This\ Way} &=& +\frac{i}{2\pi} \left(\frac{z}{L}\right)^3\left[ \frac{1}{2(z_\star-z)} +\frac{e^{-i\alpha}}{2(2z_\star-z)} + \frac{e^{-2i\alpha}}{2(3z_\star-z)} + \ldots \right] \nn\\ &=& -\frac{ie^{i\alpha}}{4\pi} \left(\frac{z}{L}\right)^3 \sum_{n=-1}^\infty \frac{e^{in\alpha}}{z+nz_\star}\nn\ee
These two series were made for each other. Summing them, and using the relationship \eqn{2d4d}, we find the 4d condensate due to the lowest Landau level given by the series
\be \plrp = -\frac{ie^{i\alpha}B}{8\pi^2} \left(\frac{z}{L}\right)^3\sum_{n=-\infty}^{+\infty}\frac{e^{in\alpha}}{z+nz_\star}\nn\ee
This series is conditionally convergent, but not absolutely convergent. This makes it interesting! As we will now show, the most interesting feature is that the sum is not continuous at $\alpha=0$. 

\para
It is a simple matter to regularise the sum by adding terms in a prescribed order. One physically motivated way of doing the sum is to first add the two terms associated to a single bounce, then the two terms associated to two bounces, and so on. Ordered in this fashion, the sum becomes,
\be F(\alpha,z) \equiv e^{i\alpha}\sum_{n=-\infty}^{+\infty}\frac{e^{in\alpha}}{z+nz_\star} = \sum_{n=0}^{+\infty}
\left[\frac{e^{in\alpha}}{z+nz_\star} +\frac{e^{-i(n+1)\alpha}}{z-(n+1)z_\star} \right]
\nn\ee
This sum $F(\alpha,z)$ is plotted in the complex plane below. Each line corresponds to the sum evaluated for a different value of $\alpha = 2\pi p/24$ with $p\in {\bf Z}$. The line itself then shows the value of $F(\alpha,z)$ as $z$ varies between $z\in [0,z_\star]$

\begin{figure}[htb]
\begin{center}
\epsfxsize=3.2in\leavevmode\epsfbox{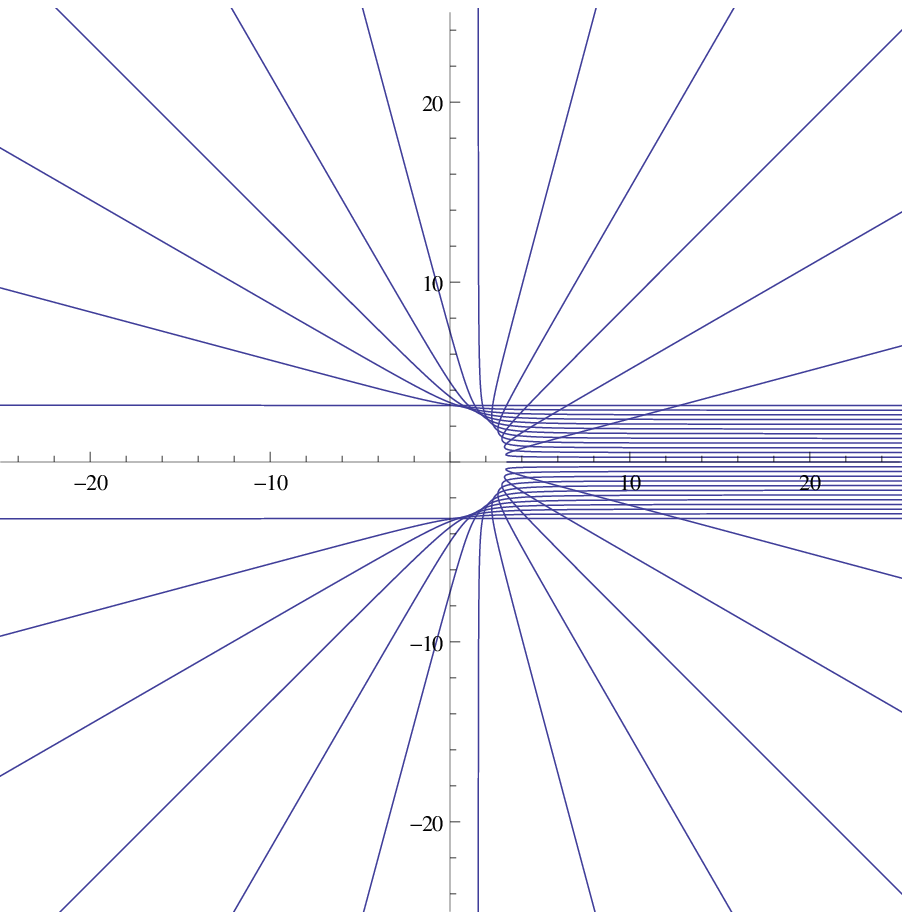}
\end{center}
\caption{Flow of the condensate in the complex plane.}\label{spiderfig}
\end{figure}

\para
Of particular interest is the gap between the two horizontal lines on the left of the plot. These two lines show the discontinuity; they correspond to $\alpha \rightarrow 0^+$ and $\alpha \rightarrow 0^-$ respectively\footnote{Numerically, these lines are plotted for $\alpha = \pm 10^{-4}\times 2\pi$ and are robust to tuning $\alpha$ to smaller values.}. As is clear from the plot, the discontinuity as $\alpha \rightarrow 0$ sits in the imaginary part of $F(\alpha, z)$. 

\para
While the numerical result above was plotted for a particular choice of regularization of the sum $F(\alpha,z)$, there is a 
more universal way of stating the main result: regardless of how you choose to perform the sum, $F(\alpha, z)$ is not single valued as $\alpha \rightarrow \alpha +2\pi$.  To determine the size of the discontinuity, we note that it arises from the log divergences in the $n\rightarrow \infty$ terms in the sum. But for large $n$, the sum can happily be approximated by the integral
\be \int_{-\infty}^{+\infty} dn\ \frac{e^{i\alpha n}}{z+nz_\star}=\frac{i\pi}{z_\star}\,{\rm sign}(\alpha) + f(z,z_\star) \nn\ee
(This equation reduces to the familiar Fourier transform for the delta function after differentiating by $\alpha$). The analytic term $f(z,z_\star)$ above differs from that of the sum since the integral is only an approximation. However, the non-analytic term ${\rm sign(\alpha)}$ has the same origin in the both the sum and integral. Indeed, it can be checked numerically that this coincides with the gap in Figure \ref{spiderfig}.

\para
Physically, this result has consequence. As $\alpha \rightarrow 0$, there remains an anomalous phase in $\plrp$ which no longer becomes purely imaginary in the limit. From our discussion in Section \ref{chiralsec}, this is interpreted as an anomalous $\pfivep$ condensate,
\be \lim_{\alpha \rightarrow 0^\pm} \pfivep= \pm \frac{1}{4\pi}\frac{BLr_\star}{r^3}\label{lovely}\ee
This is one of the main results of this paper. We will describe its implication for the dual boundary field theory in Section 4. 

\para
As should be clear from the derivation, the geometry of AdS space was not necessary for this effect.  It should also arise in a Casimir-like setting in which Dirac fermions are trapped between two, suitably reflecting, plates.

\subsection*{Discrete Symmetries in 4d}

The condensate \eqn{lovely} spontaneously breaks the discrete $CP$ transformation of the underlying theory.  Since these discrete symmetries are important for making connections with the boundary theory, we pause here to review how they work. We start with parity, $P$. In the presence of the boundary of AdS, it is convenient to define this as $x^1\rightarrow -x^1$, with $x^2$ and $r$ left unharmed. The transformation on the spinor is then
\be P: \psi \rightarrow \gamma^1\gamma^5\psi\label{parity}\ee
With this choice, $\pp$ is a scalar and parity even, while $\pfivep$ is a pseduoscalar and parity odd.

\para
With our choice of gamma matrices, charge conjugation is defined as 
\be C:\psi \rightarrow \gamma^0\gamma^2 \psi^\star\label{c}\ee
Both $\pp$ and $\pfivep$ are even under charge conjugation. (To derive this result, one must also use the Grassmann nature of $\psi$). 

\para
Finally, the magnetic field $B$ is odd under both parity and charge conjugation. The net result is that both $B$ and $\pp$ are unaffected by $CP$ and the symmetry  is spontaneously broken by $\pfivep$ as claimed.

\section{Fermion Condensates in the Boundary}\label{secbound}

We now turn to discuss the implications of the bulk fermion condensate for the dual, $d=2+1$ dimensional, boundary theory. We start by describing some simple aspects of the boundary theory.

\subsection{Aspects of the Dual Theory}

The four component Dirac spinor $\psi$ is dual to a two-component Dirac spinor operator $\Psi$ in the boundary theory. For a massless bulk fermion, the dual operator has dimension 
\be \Delta[\Psi]=\frac{3}{2}\nn\ee
This has consequence. In the large $N$ limit, the double trace operator $\bar{\Psi}\Psi$ is exactly marginal, giving rise to a line (actually, a circle) of fixed points. As explained in \cite{porrati,rr}, this can be identified with the circle of boundary conditions for the bulk spinor that we introduced in \eqn{newbound}.

\subsection*{Discrete Symmetries in 3d}

The 3d gamma matrices are inherited from their four-dimensional parents. They are
\be \gamma_{\rm 3d}^0=i\sigma^3\ \ \ , \ \ \ \gamma_{\rm 3d}^i=\sigma^i\ \ \ i=1,2
\nn\ee
Similarly, the transformations under the discrete symmetries of parity and charge conjugation are inherited from \eqn{parity} and \eqn{c} respectively. Under parity, $x^1\rightarrow -x^1$, with $x^2$ left untouched, and 
\be  P:\Psi \rightarrow \gamma^1\Psi\nn\ee
Meanwhile under charge conjugation 
\be C:\Psi \rightarrow \gamma^1\Psi^\star\nn\ee
With these definitions, we can check the charge assignments presented in the introduction: $\bar{\Psi}\Psi$ is odd under parity and even under charge conjugation. Importantly, therefore, it is odd under $CP$. 

\subsection*{Mind the Gap}

The hard wall that we introduced in Section \eqn{hardwallsec} gives rise to a gap for the ``$N^2$" degrees of freedom in the boundary theory. There are a tower of massive states which can be thought of as different bound states of the underlying large $N$ degrees of freedom. The typical  energy splitting of these states is 
\be M = \frac{r_\star}{L^2}\nn\ee
However, for the fermion spectrum with the general boundary conditions labelled by $\alpha$, there is a surprise. The gap depends on the value of $\alpha$. And, for the value $\alpha=0$ where magnetic catalysis occurs, the gap vanishes\footnote{We thank Kenny Wong for pointing this out to us.}. In other words, at $\alpha=0$ the fermions in the boundary theory are massless, despite the presence of the hard wall. This point will be discussed in more detail in a forthcoming publication \cite{bltw}.


\para
It is simple to check that the boundary conditions employed in \cite{sachdev} correspond to $\alpha =\pi$ in our notation. Here there is a gap in the fermion spectrum, but this leads to a minor puzzle since, as we described above, a mass term for a Dirac fermion in $d=2+1$ dimensions breaks parity. Yet, before we turn on a magnetic field, there is no hint of parity violation in the hard wall geometry. There is an obvious resolution to this puzzle. Parity invariant mass terms are acceptable in $d=2+1$ dimensions if Dirac fermions come in pairs with mass $\pm m$. A conserved parity operator can then be defined which acts as above while simultaneously swapping the two fermions. (Although this discussion is standard fare, a particularly nice explanation with an eye to graphene can be found in \cite{gordon}). Indeed, this is what happens in the hard wall model. If $\psi$ is an eigenfunction corresponding to a state in the boundary theory with some mass $m$ then $\gamma^r\gamma^5\psi$ is a partner solution with mass $-m$. 

\subsection{Identifying the Condensates}

In the previous section, we described the circumstances under which the bulk condensates $\pp$ and $\pfivep$ form. We would now like to understand the meaning of these in the boundary theory. It is natural to try to identify these condensates with the boundary condensate $\langle \bar{\Psi}\Psi\rangle$. But which one?

\para
The first comment to make is that these condensates arise as a one-loop effect in the bulk. (In the derivation we used only the tree-level propagator, but with the ends joined together. And a line with its ends joined together forms a loop). This means that the condensate is a ``1/N" effect in the boundary theory. Indeed, in the strict large $N$ limit all correlation functions of gauge invariant operators factorize which would imply a vanishing $\PP$. So it does seem natural that a bulk condensate and boundary condensate are related.

\para
In the introduction, we advertised the result
\be \bar{\psi}\gamma^5\psi\ \longleftrightarrow\ \bar{\Psi}\Psi\label{connect}\ee
and we will be more precise about this identification shortly. Perhaps the most compelling piece of evidence for this identification is the matching of the discrete symmetries. In particular, non-vanishing expectation values for $\pfivep$ and $\PP$ both spontaneously break CP invariance.

\para
Another argument  for the identification \eqn{connect} is to look more carefully at the way the bulk fermion encodes the source and response of the boundary operator. Near the boundary, $\psi$ has the expansion
\be 
 \psi_+ (k;r)= A(k) r^{-3/2}+ \ldots\ \ \ ,\ \ \ 
  \psi_- (k;r)= D(k)r^{-3/2} + \ldots \ \ \ \ \ \ {\rm as}\ r\rightarrow \infty
\nn\ee
where $\psi_\pm = \ft12(1\pm \gamma^r)\psi$. In the standard quantization (which means $\alpha=0$ in our previous notation), $A$ is interpreted as the source for $\Psi$, while $D$ is the response, meaning $D\sim \langle \Psi\rangle$. Now compare the two bulk bi-linears. Firstly,  $\pfivep\sim \bar{A}A + \bar{D}D$ as we approach the boundary and contains two copies of the operator. This is morally in agreement with the general framework of \cite{wittem,micha} for implementing double trace operators. In contrast, $\pp\sim \bar{A}D$ as we approach the boundary which does not look like a double trace deformation.  A similar argument was made in \cite{fermirg} when describing double trace perturbations for fermions.

\para
Finally, we will be more precise about the mapping \eqn{connect}. We introduce a composite, massless bulk scalar field $\phi = \bar{\psi}\gamma^5\psi$.  If $\phi$ is dual to $\bar{\Psi}\Psi$, then a constant $\phi$ should have the interpretation of a source for $\bar{\Psi}\Psi$. But we have already seen that this is the case: a source for $\bar{\Psi}\Psi$ corresponds to moving around the line of fixed points which, in turn, corresponds to working with a general boundary condition with $\alpha\neq 0$. But in this case, we have already seen that a constant $\phi$ condensate is formed in the bulk \eqn{plrp}.

\para
For the $\alpha=0$ boundary condition, there is no source for $\bar{\Psi}\Psi$ and, correspondingly no constant term in $\phi$. But, through the usual gauge gravity dictionary, the $1/r^3$ fall-off of $\phi$ can be  identified as the expectation value $\PP$. And, indeed, in the presence of a magnetic field, there is such a response which can be read off from \eqn{lovely}
\be \PP  \sim \frac{1}{4\pi} BM\label{done}\ee
This is our main result, converted into the boundary theory: a strongly coupled version of magnetic catalysis in $d=2+1$.

\para
We can compare this to the result to the case of free magnetic catalysis described in the introduction. In both cases, the mass of the fermion is tuned to zero. (Explicitly at weak coupling, and because we send $\alpha \rightarrow 0$ at strong coupling \cite{bltw}). And in both cases, the condensate scales linearly with the magnetic field. However, at strong coupling we see that the condensate is washed away by the presence of other massless modes. Indeed, the mass gap, $M$, for the other modes is needed to compensate for the anomalous dimension of the strongly coupled fermion. In the limit that the gap for other modes vanishes, so too does the condensate. In a generic conformal theory described by \RN black hole, no magnetic catalysis occurs.



\subsection*{What About $\pp$?}

Above we have argued that $\pfivep$ corresponds to the boundary condensate. But what about $\pp$? Does this have a signature in the boundary? The best answer that we have been able to come up with is: no. In particular, the fact that $\pp$ exists even in pure AdS tells us 
this bulk condensate is part and parcel of the vacuum CFT. 

\para
However, this feels somewhat unsatisfactory. The slogan of holography is that the boundary theory knows everything that happens in the bulk. Of course, in the presence of dynamical gravity this is manifestly true because the only diffeomorphism invariant observables are those that reside on the boundary. We leave it as an open question whether the data in $\pp$ captures some interesting physics of the boundary theory.

%
%
%

\section{Appendix: Condensate in the \RN Black Hole}
\setcounter{section}{1} \setcounter{equation}{0}
\renewcommand{\theequation}{\Alph{section}.\arabic{equation}}

In Section \ref{nobhsec}, we described the condensate arising in the presence of a magnetic field in AdS. However, we did not take into account the backreaction of the magnetic field on the geometry.  The purpose of this appendix is to describe how this changes in the black hole geometry\footnote{Landau levels for fermions in the AdS black hole were previously discussed in \cite{clifford}.}. Unfortunately, the presence of the horizon causes complication associated to the ambiguity of the vacuum state in curved space time.

\para
Allowing the  magnetic field to gravitate results in the magnetically charged \RN black hole. These solutions have constant $B$ and are parameterised by one further variable $\gamma$ which describes the deviation from extremality. The metric has the form
\be ds^2=-\frac{r^2}{L^2}f^2(r)dt^2 +\frac{L^2}{r^2}\frac{dr^2}{f^2(r)}+\frac{r^2}{L^2}(dx^2+dy^2)\nn\ee
where
\be f^2(r) = 1-\left(1+\frac{B^2}{\gamma^2r_+^4}\right)\left(\frac{r_+}{r}\right)^3+ \frac{B^2}{\gamma^2r_+^4}\left(\frac{r_+}{r}\right)^4\nn\ee
The temperature of the black hole is given by
\be T = \frac{1}{4\pi r_+}\left(3-\frac{B^2}{\gamma^2r_+^4}\right)\nn\ee
The parameter $\gamma$ takes values $\gamma^2 \geq B^2/3r_+^4$, with equality for extremal, zero temperature black holes.

\para
As in the calculation of Section \ref{nobhsec}, the magnetic field allows us to restrict to the lowest Landau level, with zero modes $\xi(r,t)$, and the action again takes the form,
\be S_{\rm eff} = \int dt dr\, \sqrt{-g}\,i\bar{\xi}\Dslash \xi\nn\ee
where the covariant derivative now feels the effect of the black hole metric, 
\be \Dslash\xi = \gamma_{\rm 2d}^0\left(\frac{1}{rf}\partial_t\xi\right) + \gamma_{\rm 2d}^1\left(rf\partial_r\xi + \frac{3}{2}f\xi+\frac{1}{2}r(\partial_r f)\xi\right)\nn\ee
Just as in AdS, this action is again equivalent to a fermion on the flat half-line. The relevant conformal scaling of the fermion is 
\be \xi_\alpha = \frac{1}{r^{3/2}f^{1/2}}\,\tilde{\xi}_\alpha\nn\ee
The action becomes that of a free fermion in terms of the spinor $\tilde{\xi}$ and the coordinate 
\be z = \frac{L^2}{r}\,g(r)\nn\ee
where the function $g(r)$ is related to $f(r)$ through the differential equation
\be r\partial_r g - g = \frac{1}{f^2}\label{whatisg}\ee
At this stage, we employ the usual flat space propagator for the rescaled spinor. Here we run into some subtleties of quantum field theory in curved space time and, in particular, with the presence of horizons. The choice of flat space propagator would seem to be akin to picking a vacuum state with respect to the time-like Killing vector $\partial_t$ outside the horizon. This is associated to the Boulware vacuum of the field in the black hole.

\para
With this choice of the propagator, the condensate again arises from the term associated to bouncing once off the boundary at $r\rightarrow \infty$ or $z=0$. 
Putting all the pieces back together, the condensate in the lowest Landau level is given by
\be \plrp = \frac{B}{4\pi}\frac{L}{r^2}\frac{1}{fg}\nn\ee
As we approach the horizon, $r\rightarrow r_+$, the function $f^2\rightarrow 0$.  For non-extremal black holes,  $f^2$ has a single zero on the horizon where we can write  $f^2\sim (r-r_+)$. Then, from \eqn{whatisg}, $g\sim \log(r-r_+)$  (since the $r\partial_rg$ term gives the leading behaviour) and we see that the condensate  diverges near the horizon. Such divergences are typical of the Boulware vacuum.

\para
For extremal black holes, $f^2$ has a double zero on the horizon. In this case, the condensate tends towards a constant as $r\rightarrow r_+$, 
\be  \plrp \rightarrow -\sqrt{6}\,\frac{B}{4\pi}\frac{L}{r_+^2}\nn\ee

\section*{Acknowledgement}
We would like to thank Ofer Aharony, Gary Horowitz, Andreas Karch, Mukund Rangamani, Gordon Semenoff and Eva Silverstein for useful discussions, and Ofer, Micha Berkooz and Shimon Yankielowicz for collaboration on related topics. We are also grateful to Elena Gubankova 
for sharing an early version of \cite{toappear} prior to publication. This research
was  supported in part by the National Science Foundation under Grant No.
PHY05-51164 and DT is grateful to KITP for hospitality while this work was completed. 
D.T. is supported by the ERC STG grant 279943, ``Strongly Coupled Systems".

\end{document}